\documentclass[final,5p,times,twocolumn]{elsarticle}

\usepackage[utf8x]{inputenc}
\usepackage[cmex10]{amsmath}
\usepackage{amsfonts}
\usepackage{amsthm}
\usepackage{algorithmic}

\usepackage{amsfonts}
\usepackage{amssymb}
\usepackage{graphicx}
\usepackage{url}
\usepackage{hyperref}
\usepackage[usenames,dvipsnames]{color}
\usepackage{caption}
\usepackage{subcaption}

\usepackage{enumitem}

\usepackage[super]{nth}
\usepackage{tikz}
\usetikzlibrary{arrows, calc,
decorations.pathmorphing,decorations.pathreplacing, decorations.markings,backgrounds, positioning,
fit, shapes.geometric, patterns}

\newtheorem{thm}{Theorem}
\newtheorem{cor}{Corollary}
\newtheorem{lem}{Lemma}
\newdefinition{defn}{Definition}
\newproof{pf}{Proof}
\newproof{rem}{Remark}

\newcommand{\veh}{P}
\newcommand{\cont}{C}
\newcommand{\ol}{M}
\newcommand{\pos}{X}
\newcommand{\totalinp}{U}
\newcommand{\inp}{W}
\newcommand{\ff}{{\mathrm{L}}}
\newcommand{\rr}{{\mathrm{R}}}
\newcommand{\contf}{\cont^\ff}
\newcommand{\contr}{\cont^\rr}
\newcommand{\olf}{\ol^\ff}
\newcommand{\olr}{\ol^\rr}
\newcommand{\inpf}{\inp^\ff}
\newcommand{\inpr}{\inp^\rr}

\newcommand{\Af}{A^\ff}
\newcommand{\Ar}{A^\rr}
\newcommand{\Bf}{B^\ff}
\newcommand{\Br}{B^\rr}
\newcommand{\indHB}{\eta}
\newcommand{\indSB}{\sigma}
\newcommand{\numInteg}{\nu}
\newcommand{\nif}{{\numInteg_{\ff}}}
\newcommand{\nir}{{\numInteg_{\rr}}}
\newcommand{\kaa}{\kappa_{\mathrm{aa}}}
\newcommand{\kbb}{\kappa_{\mathrm{bb}}}

\newcommand{\alr}{\alpha_{\rr}}
\newcommand{\alf}{\alpha_{\ff}}

\newcommand{\posRef}{\pos_{\mathrm{ref}}}

\newcommand{\extTr}{T^\rr}
\newcommand{\extTf}{T^\ff}
\newcommand{\extInputSB}{W^\rr_{\indSB}}

\begin{document}

\tikzstyle{genGraphNode} = [minimum size=3mm, thick, node
distance=5mm] 
\tikzset{
    >=stealth, bend angle=30 }
\tikzstyle{normalNode} = [genGraphNode, circle,draw=black,fill=black!10,thick]
\tikzstyle{controllingNode} = [genGraphNode,
circle,draw=black!20,fill=black!60,very thick, text=white, font=\bfseries] 
\tikzstyle{leader} = [genGraphNode,
circle,draw=black!80,fill=black!5,very thick, dashed, text=black, font=\bfseries] 
\tikzstyle{observingNode} =[genGraphNode, circle,draw=black!50,fill=white,thick]
\tikzstyle{placeholder}=[genGraphNode, circle]

\tikzstyle{block} = [draw, fill=white!20, rectangle, 
    minimum height=2em, minimum width=2em] 
\tikzstyle{blockNoDraw} = [fill=white!20, rectangle, 
    minimum height=2em, minimum width=2em] 
\tikzstyle{gain} = [draw,shape=circle, minimum height=2em, minimum
width=2em, inner sep=1pt] 
\tikzstyle{gainLeft} = [gain]
\tikzstyle{gainRight} = [gain, shape border rotate=180]
\tikzstyle{gainDown} = [gain, shape border rotate=-90]
\tikzstyle{gainUp} = [gain, shape border rotate=90]
\tikzstyle{sum} = [draw, fill=white!20, circle,inner sep=1pt]
\tikzstyle{input} = [coordinate]
\tikzstyle{output} = [coordinate]
\tikzstyle{pinstyle} = [pin edge={to-,thin,black}]

\tikzstyle{agent} = [draw, line width=3pt, rectangle, minimum height=1.3em, minimum width=1.3em, rounded corners=3pt]
\tikzstyle{agentL} = [agent, minimum height=1.7em,line width=1.8pt]
\tikzstyle{blueagent} = [agent, fill=blue!80]
\tikzstyle{redagent} = [agent, fill=red!80]
\tikzstyle{greenagent} = [agent, fill=green!80]
\tikzstyle{orangeagent} = [agent, fill=orange!80]
\tikzstyle{spring}=[thick,decorate,decoration={zigzag,pre length=0.3cm,post length=0.3cm,segment length=6}]
\tikzstyle{springShort}=[thick,decorate,decoration={zigzag,pre length=0.05cm,post length=0.05cm,segment length=3}]
\tikzstyle{thickspring}=[line width=1.8pt,decorate,decoration={zigzag,pre length=0.2cm,post length=0.2cm,segment length=6}]
\tikzstyle{redspring}=[line width=3pt,thickspring, draw=red!80]
\tikzstyle{bluespring}=[line width=3pt,thickspring, draw=blue!80]
\tikzstyle{orangespring}=[line width=3pt,thickspring, draw=orange!80]
\tikzstyle{greenspring}=[line width=3pt,thickspring, draw=green!80]

\tikzstyle{redspringShort}=[line width=1pt,springShort, draw=red!80]
\tikzstyle{bluespringShort}=[line width=1pt,springShort, draw=blue!80]
\tikzstyle{orangespringShort}=[line width=1pt,springShort, draw=orange!80]
\tikzstyle{greenspringShort}=[line width=1pt,springShort, draw=green!80]

\tikzstyle{damper}=[thick,decoration={markings,  
  mark connection node=dmp,
  mark=at position 0.5 with 
  {
    \node (dmp) [thick,inner sep=0pt,transform shape,rotate=-90,minimum width=15pt,minimum height=3pt,draw=none] {};
    \draw [thick] ($(dmp.north east)+(2pt,0)$) -- (dmp.south east) -- (dmp.south west) -- ($(dmp.north west)+(2pt,0)$);
    \draw [thick] ($(dmp.north)+(0,-5pt)$) -- ($(dmp.north)+(0,5pt)$);
  }
}, decorate]

\tikzstyle{gearbox}=[rectangle, draw, thick,decoration={markings,  
  mark=at position -0.5 with 
  {
   \draw \gear{10}{0.5}{0.8}{12}{1}{0.3};
  }
 mark=at position 5.5 with 
  {
   \draw \gear{10}{0.5}{0.8}{12}{1}{0.3};
  }
}, decorate]
\tikzstyle{ground}=[fill,pattern=north east lines,draw=none,minimum width=0.75cm,minimum height=0.3cm]

\newcommand{\gear}[6]{%
  (0:#2)
  \foreach \i [evaluate=\i as \n using {\i-1)*360/#1}] in {1,...,#1}{%
    arc (\n:\n+#4:#2) {[rounded corners=1.5pt] -- (\n+#4+#5:#3)
    arc (\n+#4+#5:\n+360/#1-#5:#3)} --  (\n+360/#1:#2)
  }%
  (0,0) circle[radius=#6] 
}

\tikzset{gradientLine/.style={
    postaction={
        decorate,
        decoration={
            markings,
            mark=at position \pgfdecoratedpathlength-0.5pt with {\arrow[blue,line width=#1] {>}; },
            mark=between positions 0 and \pgfdecoratedpathlength-8pt step 0.5pt with {
                \pgfmathsetmacro\myval{multiply(divide(
                    \pgfkeysvalueof{/pgf/decoration/mark info/distance from start}, \pgfdecoratedpathlength),100)};
                \pgfsetfillcolor{((blue)!\myval!(red)};
                \pgfpathcircle{\pgfpointorigin}{#1};
                \pgfusepath{fill};}
}}}}

\begin{frontmatter}

\title{A travelling wave approach to a multi-agent system with a path-graph topology}
\tnotetext[support]{The research was supported by the Czech Science Foundation
within the project GACR 16-19526S.}
\author[czechaddress]{Dan Martinec} 
\ead{dan.martinec@fel.cvut.cz}
\author[czechaddress]{Ivo Herman\corref{mycorrespondingauthor}}
\ead{ivo.herman@fel.cvut.cz}
\cortext[mycorrespondingauthor]{Corresponding author}
\author[czechaddress]{Michael~\v{S}ebek}
\ead{sebekm1@fel.cvut.cz}
\address[czechaddress]{Faculty of Electrical Engineering, Czech Technical
University in Prague. Czech Republic}



\begin{abstract}
The paper presents a novel approach for the analysis and control of a multi-agent system with non-identical agents and a path-graph topology. With the help of irrational wave transfer functions, the approach describes the interaction among the agents from the `local' perspective and identifies travelling waves in the system. It is shown that different dynamics of the agents creates a virtual boundary that causes a partial reflection of the travelling waves. Undesired effects due to the reflection of the waves, such as amplification/attenuation, long transients or string instability, can be compensated by the feedback controllers introduced in this paper. We show that the controllers achieve asymptotic and even string stability of the system.
\end{abstract}

\end{frontmatter}


\section{Introduction}
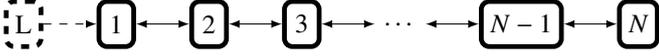
\begin{figure}[t]
	\begin{tikzpicture}[auto, >=latex, node distance = 0.7cm]
	\node[agentL] (n1) { $1$};
	\node[agentL] (n2) [right=of n1]{$2$}
		edge [<->](n1);
	\node[agentL, dashed] (L) [left=of n1] {L}
		edge [->, dashed] (n1); 
	\node[agentL] (n3) [right=of n2]{$3$}
		edge [<->](n2);
	\node (dots) [right=of n3] {\ldots}
		edge [<->] (n3);
	\node[agentL] (nn1) [right=of dots]{$N-1$}
		edge [<->](dots);
	\node[agentL] (nn) [right=of nn1]{$N$}
		edge [<->](nn1);
	
\end{tikzpicture}
	\caption{Structure of the path graph. The arrows show interaction between agents. The agent $L$ is a virtual leader of the multi-agent system, which commands only the first agent.}
	\label{fig:pathGraph}
\end{figure}

A path graph is one of the simplest and most studied interaction topologies of a multi-agent system.  This topology is used in many applications, such as vehicle platoons \cite{Lin2012, Knorn2014}, discretized flexible structures \cite{Singhose1996,Dwivedy2006}, or a spatially-discretized models of long electrical transmission lines \cite{Dhaene1992}. Formally, a path graph is a graph with $N$ vertices, ordered as $v_1, v_2, \ldots, v_N$, with edges between vertices $\{v_i, v_{i+1}\},\,\, i=1, \ldots, N-1$. Equivalently, in a path graph topology, each agent, except for the first and last one, interacts with its two neighbours (see Fig. \ref{fig:pathGraph}).

There are many tools for describing a multi-agent systems. They range from state-space techniques \cite{Olfati-Saber2007}, polynomial approaches \cite{Hara2014} to statistical-physics-based description \cite{Koorehdavoudi2016}. In path graphs, scaling of $\mathcal{H}_2$ and $\mathcal{H}_\infty$ norms was calculated in \cite{Lin2012} and \cite{Herman2014c}. Boundedness of the norm of interest for any $N$ can be captured by a term ``string stability''. Roughly speaking, in a string-stable system a disturbance is not amplified as it propagates among the agents (see \cite{Ploeg2014} for various definitions). The approaches mentioned so far are very useful in analyzing aggregate properties of the multi-agent system, such as its stability or system norms. On the other hand, it is more difficult to infer from them what happens in the middle of the system or near the boundaries. 

Following the ideas of other researchers (e.g., \cite{Barooah2009, Galbusera2007}), we will describe the system using a wave perspective. 
Indeed, the propagation of the change in the multi-agent system can be described with the help of travelling waves. We will illustrate it on an example of a system with identical agents and a path-graph topology. If the first agent changes its output, then all following agents sequentially respond to this change. If we study their response from the local point of view \cite{OConnor2007,Martinec2014a,Sirota2015} we can notice that the change is propagated as a wave. The wave departs from the first agent and travels along the system to the last agent, where it reflects and travels back. When it reaches the first agent, it reflects again. A similar phenomenon is apparent when the agents are non-identical---the travelling wave is partially reflected on non-identical agents \cite{Martinec2014b}. We can imagine this behaviour as the reflection of the wave if it encounters a boundary between two media of different properties. 

The tool for analysis in this paper will be so called \emph{wave transfer function} (WTF). The transfer-function approach to waves has recently been revisited in a series of papers for lumped models \cite{OConnor2007,Yamamoto2016} and for continuous flexible structures \cite{Sirota2015}. The travelling wave approach has also been applied to vibration control \cite{Mei2011} and it seems to be related to the impedance matching in the power networks \cite{Lesieutre2002}. The wave-based description leads to irrational transfer functions, analysis of which differs in several aspects from their rational counterparts \cite{Curtain2009}. 

This paper continues in the research started in \cite{Martinec2014a}, where waves in a platoon of identical vehicles were considered. A natural extension of this model is to consider a chain of non-identical (heterogeneous) agents. The first step in the treatment of non-identical agents using travelling waves is given in \cite{OConnor2011} for a mass-spring model. We generalize it by considering arbitrary dynamics of the agents and their controllers. The preliminary results are presented in \cite{Martinec2014b}, where we introduced the \emph{soft boundary} in a chain of vehicles. Here, we introduce the second fundamental type of boundary, the \emph{hard boundary}. Although the boundaries are virtual in nature, they principally affect the overall system behaviour. We present some fundamental properties of the boundaries and design wave-absorbing controllers for both types of boundaries.

The main contributions of the paper are: i) mathematical description of the travelling waves in a multi-agent system with non-identical agents given by Theorems \ref{soft_boundary_theorem} and \ref{hard_boundary_theorem}, ii) a design of a controller that prevents a reflection of the travelling wave (Theorems \ref{lem:soft_boundary_control} and \ref{lem:hard_boundary_control}) and iii) proof of stability and string stability when these controllers are used (Theorem \ref{lem:soft_boundary_stability}). For better understanding and easy simulations, we provide a set of functions in MATLAB, see \emph{WaveBox} \cite{WaveBox2015}.

\section{System model}
In the whole paper we work only with LTI systems in the Laplace domain, all transfer functions are SISO and signals are assumed to be scalars. The argument $(s)$ denotes the Laplace variable and can be omitted when no ambiguity seems possible.

\begin{figure}[t]
 \centering
  \begin{tikzpicture}[auto, >=latex]
	\node[block] (P) { $\veh_{i}(s)$};
	\node[sum] (sum1) [left=of P,left=2em] {$+$};
	\draw [->]	(sum1) -- (P) node[above, midway] {$\totalinp_i$};
	\node [sum, fill=black, minimum width=0.01em, minimum height=0.01em] (p1) [right= 1.8em of P] {};
	\draw [->] (P) -- (p1) node[midway, above] { $\pos_i$};
	\draw [->] (p1.east) -- ([xshift=1.5em]p1.east);
	
	\node[block] (Cf) [above left= 0.01em and 1em of sum1] {$\contf_i(s)$};
	\draw [->] (Cf) -| (sum1);
	\node[block] (Cr) [below left= 0.01em and 1em of sum1] {$\contr_i(s)$};
	\draw [->] (Cr) -| (sum1);
	\node[sum] (sumr1) [left= 0.5em of Cr] {\tiny{$+$}}
		edge [->] (Cr);
	\node[sum] (sumr2) [left= 0.5em of sumr1] {\tiny{$+$}}
		edge [->] (sumr1);
	\node[sum] (sumf1) [left= 0.5em of Cf] {\tiny{$+$}}
		edge [->] (Cf);
	\node[sum] (sumf2) [left= 0.5em of sumf1] {\tiny{$+$}}
		edge [->] (sumf1);
	\draw [->] ([xshift=-3em] sumf2.west) -- (sumf2.west) node[near start, above] {$\pos_{i-1}$};
	\draw [->] ([xshift=-3em] sumr2.west) -- (sumr2.west) node[near start, above] {$\pos_{i+1}$};
	\draw [->] ([yshift=+1.8em, xshift=-3.40em] sumf2.north) -| (sumf2.north) node[at start, above, xshift=0.5em] {$\inpf_i$};
	\draw [->] ([yshift=-1.8em, xshift=-3.40em] sumr2.south) -| (sumr2.south) node[at start, above, xshift=0.5em] {$\inpr_i$};

	\draw[<-] (sumf1.north) -- ++(0.0em,2.5em)node[near start, right, xshift=-0.15em, yshift=-0.2em] {\footnotesize$-$} -| (p1.north) ;
	\draw[<-] (sumr1.south) -- ++(0.0em,-2.5em)node[near start, xshift=-0.15em, yshift=+0.2em] {\footnotesize$-$} -| (p1.south) ;
	\draw [decorate,decoration={brace,amplitude=10pt,mirror,raise=1pt}, dashed] (P.north east) -- (Cf.north west) node [midway,xshift=+0.6cm, yshift=+0.9cm] { $\olf_{i}(s)$};
	\draw [decorate,decoration={brace,amplitude=10pt,raise=1pt}, dashed] (P.south east) -- (Cr.south west) node [midway,xshift=-0.4cm, yshift=-0.3cm] { $\olr_{i}(s)$};
	
	\draw [dashed] ([yshift=-3em, xshift=-1em] sumr2.south) rectangle ([yshift=+4.8em, xshift=+0.3em] p1.east);
\end{tikzpicture}
  \caption{The model of $i$th agent.}
  \label{fig:agent_model}
\end{figure}
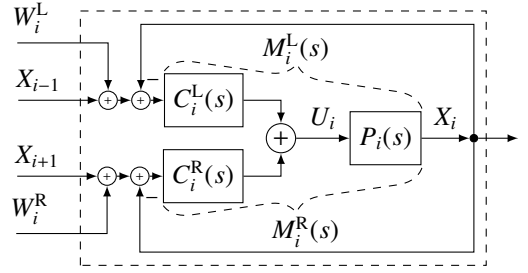

We consider a multi-agent system of $N$ non-identical agents with a path-graph interaction topology. Each agent interacts with its nearest neighbors using output feedback. The LTI dynamics of the agent $i$ consists of two parts: the model of the plant $\veh_i(s)$ and the models of the controllers. The controller $\contf_i(s)$ processes the output error $\pos_{i-1}(s)-\pos_{i}(s)$ to agent's predecessor $i-1$ and the controller $\contr_i(s)$ processes the output error $\pos_{i+1}(s)-\pos_{i}(s)$ to agent's follower $i+1$. Based on the order of agents in Fig. \ref{fig:pathGraph}, we denote the interaction of the agent $i$ with agent the $i-1$ by superscript 'L' (\emph{left} of agent $i$) and the interaction with agent $i+1$ by superscript 'R' (\emph{right} of agent $i$). The input $\totalinp_i(s)$ of the plant is generated by the controllers as
\begin{equation}
\begin{aligned}
	\totalinp_i(s) =& \contf_i(s)\left(\pos_{i-1}(s)-\pos_i(s)+\inpf_i(s)\right) \\
	&+ \contr_i(s)\left(\pos_{i+1}(s)-\pos_i(s)+\inpr_i(s)\right),
	\end{aligned}
\end{equation}
where $\inpf_i(s)$ and $\inpr_i(s)$ are external inputs to the agent, which will be defined later on.
The output of the agent is then given by $\pos_i(s)=\veh_i(s) \totalinp_i(s)$. By defining the \emph{left} open-loop transfer function (OLTF) $\olf_{i}(s) =  \veh_{i}(s) \contf_i(s)$ and \emph{right} OLTF $\olr_{i}(s) =  \veh_i(s) \contr_i(s)$, we obtain the overall model of the agent
\begin{equation}
\begin{aligned}
 \pos_{i}(s) =& \olf_{i}(s) \Big(\pos_{i-1}(s)-\pos_{i}(s) \Big) + \olr_{i}(s)\Big(\pos_{i+1}(s)-\pos_{i}(s)\Big)\\
 &+ \olf_{i}(s)\inpf_i(s) + \olr_{i}(s)\inpr_i(s).
\end{aligned}\label{eq:eq1}
\end{equation}
The structure of the $i$th agent is depicted in Fig.~\ref{fig:agent_model}. Usually, there is at least one integrator both in the left and right OLTFs (for instance, from velocity to position), such that the OLTF can be factored as $\ol(s)=1/s^\numInteg \overline{\ol}(s)$, where $\overline{\ol}(0)<\infty$ and $\numInteg$ is the number of integrators in the corresponding open loop (either $\olf_{i}$ or $\olr_{i}$). 

Often, the system has a reference $\posRef$ which it should track. This reference is given by a (virtual) leader, which is usually connected to one of the end nodes. Without loss of generality, we assume that it is connected to the first agent, which is then described as
  $\pos_1(s) = \olf_{1}(s)(\posRef(s)-\pos_{1}(s)+\inpf_1(s)) + \olr_{1}(s)(\pos_{2}(s)-\pos_{1}(s)+\inpr_1(s))$,
The last agent $i=N$ has has only one neighbor, so its model is $
  \pos_{N}(s) = \olf_{N}(s)\big(\pos_{N-1}(s)-\pos_{N}(s)+\inpf_N(s)\big).$

\subsection{Wave transfer function}
The key idea of the wave approach is that the output of the $i$th agent is decomposed into two components, $A_i(s)$ and $B_i(s)$ such that $X_i(s)=A_i(s)+B_i(s)$. The component $A_i(s)$ represents a wave which propagates from left to right, that is, to the agents with higher indices. The component $B_i(s)$ represents the wave propagating from right to left---to the agents with lower indices. The idea is similar to the standard D'Alambert solution of the wave equation in PDE, where also two waves propagating in different directions appear \cite{french2003}. 

Now we summarize the results of \cite{Martinec2014a}, where we considered identical agents. In this case $\olf_i(s) = \olr_{i}(s) = M(s)$.
The \emph{Wave transfer function} $G(s)$ captures how the wave propagates in the system in one direction, that is $A_{i+1}(s)=G(s) A_i(s)$ and $B_{i-1}(s)=G(s) B_i(s)$. There is a simple way how to derive this transfer function. Consider a path graph with infinite number of agents and with only a wave propagating in the direction of increasing index. Since there is no end in this system, the wave will never reflect back, so $X_i(s)=A_i(s)$. Then
WTF is given by $G(s) = X_{i+1}(s)/X_{i}(s)$ for $N \rightarrow \infty$, see \cite[Sec. 3.1]{Martinec2014a}. When $\inpf_{i}\!=\!\inpr_{i}\!=\!0$, the system with identical agents is described for $i \in [1, N-1]$ as 
\begin{align}
  X_i(s) &= A_i(s) + B_i(s),
  \label{eq:pos_decomp}\\
  A_{i+1}(s) &= G(s)A_i(s),
  \label{eq:anp1}\\
  B_{i}(s) &= G(s)B_{i+1}(s),
  \label{eq:bnp1}\\
	G(s)&= \frac{1}{2}\alpha(s) - \frac{1}{2}\sqrt{\alpha^2(s)-4},
	  \label{eq:wtf_intro1}
\end{align}
where $\alpha(s) = 2+ 1/M(s)$, or, alternatively, $\alpha(s) = G(s)+G^{-1}(s)$. The function $G^{-1}(s) = 1/G(s) = \!\frac{1}{2}\alpha(s)\! +\! \frac{1}{2}\sqrt{\alpha^2(s)\!-\!4}$. We now explain the traveling wave concept. Combining (\ref{eq:pos_decomp})-(\ref{eq:bnp1}), 
\begin{equation}
X_i(s) = G(s) A_{i-1}(s) + G(s) B_{i+1}(s). \label{eq:waveProp}
\end{equation}
This means that the wave $A_{i-1}(s)$ coming from the left (from the agent with lower index) is transformed through the transfer function $G(s)$ and summed with the transformed wave $G(s) B_{i+1}(s)$ from the right (from the agent with higher index). 

\begin{figure}[t]
 \centering
  \centering
	\begin{tikzpicture}[auto, >=latex]
	\node (P1) {$X_{\mathrm{ref}}$};
	\node[agentL]  (N1)[right=of P1, right=1.3em] {$X_1$};	
	\node[agentL]  (Nim1)	[right=of N1, right=2em] {$X_{i-1}$}; 
	\node[agentL] (Ni) [right=of Nim1, right=3em] {$X_i$};
	\node[agentL] (Nip1) [right=of Ni, right=3em] {$X_{i+1}$};
	\node[agentL] (NN) [right=of Nip1, right=2em] {$X_N$};

	\node (T1b) [above=of N1, above=0.3em] {$B_{1}$};
	\node (T1a) [below=of N1, below=0.3em] {$A_{1}$};
	\node (Tim1b) [above=of Nim1, above=0.3em] {$B_{i-1}$};
	\node (Tim1a) [below=of Nim1, below=0.3em] {$A_{i-1}$};
	\node (Tib) [above=of Ni, above=0.3em] {$B_{i}$};
	\node (Tia) [below=of Ni, below=0.3em] {$A_{i}$};
	\node (Tip1b) [above=of Nip1, above=0.3em] {$B_{i+1}$};
	\node (Tip1a) [below=of Nip1, below=0.3em] {$A_{i+1}$};
	\node (TNb) [above=of NN, above=0.3em] {$B_{N}$};
	\node (TNa) [below=of NN, below=0.3em] {$A_{N}$};
	
	\node (Tm1b) [right=of T1b, right=0.3em] {};
	\node (Tm1a) [right=of T1a, right=0.3em] {};
	\node (Tmb) [left=of Tim1b, left=0.3em] {};
	\node (Tma) [left=of Tim1a, left=0.3em] {};
	
	\node (Tp1b) [right=of Tip1b, right=0.3em] {};
	\node (Tp1a) [right=of Tip1a, right=0.3em] {};
	\node (Tpb) [left=of TNb, left=0.3em] {};
	\node (Tpa) [left=of TNa, left=0.3em] {};
	
	\draw [->] (P1) -- (N1);
	\draw [line width=3pt, line cap=round, dash pattern=on 0pt off 2.8\pgflinewidth, draw] (N1) -- (Nim1);
	\draw [thickspring] (Nim1) -- (Ni);
	\draw [thickspring] (Ni) -- (Nip1);
	\draw [line width=3pt, line cap=round, dash pattern=on 0pt off 2.8\pgflinewidth, draw] (Nip1) -- (NN);
	
	\draw [->] (T1a) -- (Tm1a.center);
	\draw [->] (Tma.center) -- (Tim1a);
	\draw [->] (Tim1a) to node [midway, below] {$G$} (Tia);
	\draw [->] (Tia) to node [midway, below] {$G$} (Tip1a);
	\draw [->] (Tip1a) -- (Tp1a.center);
	\draw [->] (Tpa.center) -- (TNa);
	\path (TNa.+0) edge [bend right, ->]node [midway, right] {$G$} (TNb.+0) ;
	
	\draw [->] (TNb) -- (Tpb.center);
	\draw [->] (Tp1b.center) -- (Tip1b);
	\draw [->] (Tip1b) to node [midway, above] {$G$} (Tib);
	\draw [->] (Tib) to node [midway, above] {$G$} (Tim1b);
	\draw [->] (Tim1b) -- (Tmb.center);
	\draw [->] (Tm1b.center) -- (T1b);
	\path (T1b.+180) edge [bend right, ->] node [near start, left, yshift=0.5em, xshift=0.3em] {$-G^2$}  (T1a.-180) ;
\end{tikzpicture}

%
  \caption{Schematic of a multi-agent system with identical agents. The squares are agents with local dynamics described by (\ref{eq:eq1}). The virtual connections, created by the local control laws, are illustrated by springs. }
  \label{fig:homogenous_path_graph}
\end{figure}
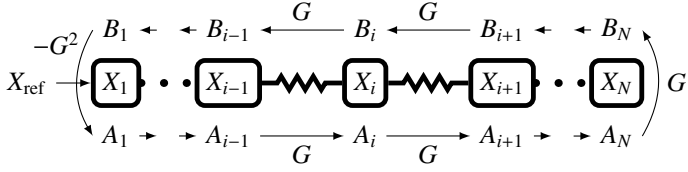

Now suppose that the number of agents is finite. Then, as we discussed in \cite{Martinec2014a}, there are two types boundaries in the homogeneous system, located at the end nodes in the path graph. The \emph{forced-end boundary} is caused by the leader's output $\posRef$. If the leader affects the first agent, the boundary is described by
\begin{align}
  A_1(s) = G(s)\posRef(s) - G^2(s) B_1(s).\label{eq:forced_end}
\end{align}
The \emph{free-end boundary} is at the end node which has only one neighbor. If it is located at the $N$th agent, it is given by
\begin{align}
  B_N(s) = G(s) A_N(s).\label{eq:free_end}
\end{align}
The first type of the boundary is analogous to Dirichlet boundary condition (``zero position'') and the second to Neumann (``zero derivative with respect to position") \cite{french2003}. The signal propagation in the system with boundaries is shown in Fig.~\ref{fig:homogenous_path_graph}. 


\subsection{Soft and hard boundaries}
Now let us go back to the heterogeneous system of non-identical agents. We define additional boundaries between the agents having different models. In general, we can distinguish between three cases: i) $\olr_{i}(s) \neq \olf_{i+1}(s)$, ii) $\olf_{i}(s) \neq \olr_{i}(s)$, and iii) a combination of i) and ii). As we will see, these boundaries cause a partial reflection of the travelling wave. In order to derive their mathematical properties, we assume that the boundary of interest is the \emph{only boundary} in the system. This also means that all the agents to the left from the boundary are identical. Let the WTF corresponding to the open loop for the agents to the left be $G(s)$. Similarly, the agents to the right of the boundary are identical and the WTF for them is $H(s)$. 

We will focus on the boundaries caused by i) and ii). Once their properties are known, the description of their combination iii) is easy. Thus, we do not discuss it in this paper. Instead, we refer the interested reader to the dissertation \cite{Martinec2015d}.
\begin{defn}
  \emph{Soft boundary} is a virtual boundary \emph{between} two agents, indexed $\indSB$ and $\indSB+1$, with the  property $\olr_{\indSB}(s) \neq \olf_{\indSB+1}(s).$
\end{defn}
The soft boundary is, for instance, located in a platoon of non-identical vehicles governed by the same symmetric bidirectional control law \cite{Knorn2014} or in a mass-spring model with identical springs but non-identical masses. The wave transfer function between the agents with index $i \leq \indSB$ is $G(s)$ and between the agents with $i\geq\indSB+1$ is $H(s)$. Using the same approach as in \cite[Sec. 3.1]{Martinec2014a}, these wave transfer functions are given as
\begin{align}
    G(s) = \frac{1}{2}\alpha_1-\frac{1}{2}\sqrt{\alpha_1^2-4},\;\;\; H(s) = \frac{1}{2}\alpha_2-\sqrt{\alpha_2^2-4},
    \label{eq:sb_3}
  \end{align}
where $\alpha_1 = 2 + 1/\olr_{\indSB}(s)$ and $\alpha_2 = 2+1/\olf_{\indSB+1}(s)$. For the agents with indices $i \leq \indSB$ holds $\olf_{i}=\olr_{i}=\olr_{\indSB}$ and for agents $i \geq \indSB+1$ holds $\olf_{i}=\olr_{i}=\olf_{\indSB+1}$. 

The second type of boundary is defined as follows.
\begin{defn}
  The \emph{hard boundary} is a virtual boundary located \emph{at} the $\indHB$th agent with the property    $\olf_{\indHB} (s) \neq \olr_{\indHB} (s).$
\end{defn}
The hard boundary is, for instance, located in a platoon of identical vehicles with asymmetric control, see \cite{Barooah2009}, or in a mass-spring model with identical masses but different springs.

\begin{figure}[t]
	\centering
	\begin{tikzpicture}[auto, >=latex]
	\node[placeholder] (P1) {};
	\node[blueagent, fill=blue!80]  (N1)[right=of P1, right=1cm] {};	
	\node[blueagent]  (N2)	[right=of N1, right=3em] {}; 
	\node[placeholder] (P2) [right=of N2, right=1.0em] {};
	\node[redagent]  (N3)	[right=of P2, right=1.0em] {}; 
	\node[redagent]  (N4)	[right=of N3, right=3em] {}; 
	\node (P3) [right=of N4, right=1cm] {};
	
	\node (T1b) [above=of N1, above=0.3em] {$B_{\indSB-1}$};
	\node (T1a) [below=of N1, below=0.3em] {$A_{\indSB-1}$};
	\node (T2b) [above=of N2, above=0.3em] {$B_{\indSB}$};
	\node (T2a) [below=of N2, below=0.3em] {$A_{\indSB}$};
	\node (T3b) [above=of N3, above=0.3em] {$B_{\indSB+1}$};
	\node (T3a) [below=of N3, below=0.3em] {$A_{\indSB+1}$};
	\node (T4b) [above=of N4, above=0.3em] {$B_{\indSB+2}$};
	\node (T4a) [below=of N4, below=0.3em] {$A_{\indSB+2}$};
	\node (T0b) [left=of T1b, left=3em] {};
	\node (T0a) [left=of T1a, left=3em] {};
	\node (T5b) [right=of T4b, right=3em] {};
	\node (T5a) [right=of T4a, right=3em] {};
	\draw [line width=3pt, draw=blue!80, line cap=round, dash pattern=on 0pt off 3\pgflinewidth] (P1) -- (N1);
	\draw [bluespring] (N1) -- (N2);
	\draw [line width=3pt,decorate,decoration={zigzag,pre length=0.2cm,segment length=6}, draw=blue!80] (N2) -- (P2.center);
	\draw [line width=3pt,decorate,decoration={zigzag,pre length=0.2cm,post length=0.0cm, segment length=6}, draw=red!80] (N3) -- (P2.center);
	\draw [redspring] (N3) -- (N4);
	\draw [line width=3pt, line cap=round, dash pattern=on 0pt off 3\pgflinewidth, draw=red!80] (N4) -- (P3);
	
	\draw [->] (T1b) -- (T0b);
	\draw [->] (T2b) -- (T1b);
	\path (T3b) edge [bend right, ->] (T2b);
	\draw [->] (T4b) -- (T3b);
	\draw [<-] (T4b) -- (T5b);
	
	\draw [<-] (T1a) -- (T0a);
	\draw [<-] (T2a) -- (T1a);
	\path (T3a) edge [bend left, <-] (T2a);
	\draw [<-] (T4a) -- (T3a);
	\draw [->] (T4a) -- (T5a);
	
	\path (T2a.+0) edge [bend right, bend angle=45, ->] (T2b.+0);
	\path (T3a.-180) edge [bend left,bend angle=45, <-] (T3b.-180);
\end{tikzpicture}\\
	\begin{tikzpicture}[auto, >=latex]
	\node[placeholder] (P1) {};
	\node[blueagent, fill=blue!80]  (N1)[right=of P1, right=1em] {};	
	\node[blueagent]  (N2)	[right=of N1, right=3em] {}; 
	\node[agent, shading = axis,left color=blue!80, right color=red!80]  (N3)	[right=of N2, right=3em] {}; 
	\node[redagent]  (N4)	[right=of N3, right=3em] {}; 
	\node[redagent]  (N5)	[right=of N4, right=3em] {}; 
	\node (P3) [right=of N5, right=1em] {};
	
	\node (T1b) [above=of N1, above=0.3em] {$B_{\indHB-2}$};
	\node (T1a) [below=of N1, below=0.3em] {$A_{\indHB-2}$};
	\node (T2b) [above=of N2, above=0.3em] {$B_{\indHB-1}$};
	\node (T2a) [below=of N2, below=0.3em] {$A_{\indHB-1}$};
	\node (T3la) [right=of T2a, right=0.6em] {$\Af_{\indHB}$};
	\node (T3lb) [right=of T2b, right=0.6em] {$\Bf_{\indHB}$};
	\node (T4b) [above=of N4, above=0.3em] {$B_{\indHB+1}$};
	\node (T4a) [below=of N4, below=0.3em] {$A_{\indHB+1}$};
	\node (T3ra) [left=of T4a, left=0.6em] {$\Ar_{\indHB}$};
	\node (T3rb) [left=of T4b, left=0.6em] {$\Br_{\indHB}$};	
	\node (T5b) [above=of N5, above=0.3em] {$B_{\indHB+2}$};
	\node (T5a) [below=of N5, below=0.3em] {$A_{\indHB+2}$};
	\node (T0b) [left=of T1b, left=1em] {};
	\node (T0a) [left=of T1a, left=1em] {};
	\node (T6b) [right=of T5b, right=1em] {};
	\node (T6a) [right=of T5a, right=1em] {};
	\draw [line width=3pt, draw=blue!80, line cap=round, dash pattern=on 0pt off 3\pgflinewidth] (P1) -- (N1);
	\draw [bluespring] (N1) -- (N2);
	\draw [bluespring] (N2) -- (N3);
	\draw [redspring] (N3) -- (N4);
	\draw [redspring] (N4) -- (N5);
	\draw [line width=3pt, line cap=round, dash pattern=on 0pt off 3\pgflinewidth, draw=red!80] (N4) -- (P3);
	
	\draw [->] (T1b) -- (T0b);
	\draw [->] (T2b) -- (T1b);
	\draw [->] (T3lb) -- (T2b);
	\draw [->] (T3rb) -- (T3lb);
	\draw [->] (T4b) -- (T3rb);
	\draw [->] (T5b) -- (T4b);
	\draw [<-] (T5b) -- (T6b);
	
	\draw [<-] (T1a) -- (T0a);
	\draw [<-] (T2a) -- (T1a);
	\draw [<-] (T3la) -- (T2a);
	\draw [<-] (T3ra) -- (T3la);
	\draw [<-] (T4a) -- (T3ra);
	\draw [<-] (T5a) -- (T4a);
	\draw [<-] (T6a) -- (T5a);
	
	\draw [->] (T3la.+70) --  (T3lb.-70);
	\draw [<-] (T3ra.+110) -- (T3rb.-110);

\end{tikzpicture}
	\caption{A system with the soft (top) and hard (bottom) boundary. The blue and red squares are agents with the WTFs $G(s)$ and $H(s)$ in (\ref{eq:sb_3}), (\ref{eq:asymmetric5}). The blue-red spring is the soft boundary and the blue-red square is the hard boundary.}
  \label{fig:boundaries_soft_tf}
\end{figure}
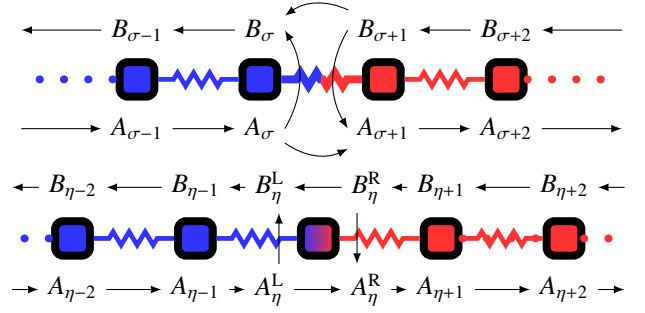

The adjective `hard' emphasizes the fact that the boundary is located at an agent, in contrast to the soft boundary, located between two agents. To distinguish between the incident, transmitted and reflected waves at the hard boundary, we decompose $X_{\indHB}$ to the hard-boundary wave components as
\begin{align}
  X_{\indHB}(s) &= \Af_{\indHB}(s) + \Bf_{\indHB}(s) = \Ar_{\indHB}(s) + \Br_{\indHB}(s),
  \label{eq:asymmetric2}
  \end{align}
where the indexes $\text{L}$ and $\text{R}$ denote the wave components that are next to the left and right sides of the boundary, respectively. They are tied by the following.
\begin{align}
\Af_\indHB(s) &= G(s) A_{\indHB-1}(s), \;\;\; \Bf_\indHB(s) = G^{-1}(s) B_{\indHB-1}(s), \label{eq:asymmetric3}\\
 \Ar_{\indHB}(s) &= H^{-1}(s) A_{\indHB+1}(s), \; \Br_{\indHB}(s) = H(s) B_{\indHB+1}(s),
 \label{eq:asymmetric4}
\end{align}
where the WTFs to the left ($G(s)$) and right ($H(s)$) of the boundary are using the approach in \cite[Sec. 3.1]{Martinec2014a} given as
  \begin{align}
    G(s) = \frac{1}{2}\alpha_1-\frac{1}{2}\sqrt{\alpha_1^2-4},\;\;\; H(s) = \frac{1}{2}\alpha_2-\sqrt{\alpha_2^2-4},\label{eq:asymmetric5}
  \end{align}
and $\alpha_1 = 2 + 1/\olf_{\indHB}(s)$ and $\alpha_2 = 2+ 1/\olr_{\indHB}(s)$.
Both boundaries are illustrated in Fig.~\ref{fig:boundaries_soft_tf}. 
\vspace{-10pt}
\section{Mathematical description of waves in path graph}
In this section we will concentrate on the description of the boundaries and their effect on the propagating wave. 
\begin{thm}[\cite{Martinec2014b}]
\label{soft_boundary_theorem}
  Let $G(s)$ and $H(s)$ be given by (\ref{eq:sb_3}). Then a soft boundary is in the Laplace domain described by four boundary-transfer functions (BTFs)
  \begin{alignat}{2}
    T_{\textnormal{aa}} &= \frac{A_{\indSB+1}}{A_{\indSB}} =  \frac{H-H G^2}{1-HG}, \;\;\; T_{\textnormal{ba}}\, &=& \,\frac{A_{\indSB+1}}{B_{\indSB+1}}= \frac{H G - H^2}{1-HG},\label{eq:sb_1}\\
    T_{\textnormal{bb}} &= \frac{B_{\indSB}}{B_{\indSB+1}} = \frac{G - H^2 G}{1-HG}, \;\;\; T_{\textnormal{ab}}\, &=&\, \frac{B_{\indSB}}{A_{\indSB}}= \frac{HG - G^2}{1-HG}, \label{eq:sb_2}
  \end{alignat}
  
\end{thm}

The proof is given in \cite{Martinec2014b}. The interpretation is that if there is a wave travelling to the soft boundary from the left, then it is partially reflected from the boundary (described by $T_{\text{ab}}$) and partially transmitted through the boundary (by $T_{\text{aa}}$). Likewise, if the wave travels from the right, then the transfer functions $T_{\text{ba}}$ and $T_{\text{bb}}$ represent the respective waves. When combined,
\begin{align}
  X_{\indSB}(s) &= G(s) (1+ T_{\text{ab}}(s)) A_{\indSB-1}(s) + T_{\text{bb}}(s) B_{\indSB+1}(s), \label{eq:SB_4a}\\
  X_{\indSB+1}(s) &= H(s) (1+ T_{\text{ba}}(s)) B_{\indSB+2}(s) + T_{\text{aa}}(s) A_{\indSB}(s).\label{eq:SB_4b}
\end{align}
The forced-end boundary is an example of the soft boundary. Substituting $G=0$ into (\ref{eq:sb_1}) and (\ref{eq:sb_2}) gives $T_{\text{aa}} = H$, $T_{\text{ba}} = -H^2$ and $T_{\text{bb}} = T_{\text{ab}} = 0$. In addition, if $G(s)=H(s)$ (there is no boundary), there there is no reflection ($T_\text{ab}(s)=T_\text{ba}(s)=0$) and everything gets through ($T_\text{aa}(s)=G(s), T_\text{bb}(s)=G(s)$).

\begin{thm}
\label{hard_boundary_theorem}
  Let $G$ and $H$ be given by (\ref{eq:asymmetric5}). Then the BTFs describing the hard boundary in the Laplace domain are
  \begin{align}
  T_{\textnormal{AA}}\! &=\! \frac{\Ar_{\indHB}}{\Af_{\indHB}}\!=\!\frac{(1+G)(1-H)}{1-HG}, \;\;\;\;
T_{\textnormal{BA}}\!=\! \frac{\Ar_{\indHB}}{\Br_{\indHB}}\! =\! \frac{H-G}{1-HG},\label{eq:HB_1a}\\
  T_{\textnormal{BB}}\! &=\! \frac{\Bf_{\indHB}}{\Br_{\indHB}}\! =\! \frac{(1+H)(1-G)}{1-HG}, \;\;\;\;
T_{\textnormal{AB}}\! =\! \frac{\Bf_{\indHB}}{\Af_{\indHB}}\! =\! \frac{G-H}{1-HG},\label{eq:HB_1b}
\end{align}

\end{thm}
\begin{pf}
  From (\ref{eq:eq1}), the output of the agent can be rewritten as
\begin{align}
  X_{\indHB}(s) = T_{\text{L}}(s) X_{\indHB-1}(s) + T_{\text{R}}(s) X_{\indHB+1}(s),
  \label{eq:app_hard_boundary1}
\end{align}
where $T_{\text{L}} = \olf_{\indHB}/(1+\olf_{\indHB} + \olr_{\indHB})$ and $T_{\text{R}} = \olr_{\indHB}/(1+\olf_{\indHB} + \olr_{\indHB})$. We combine (\ref{eq:pos_decomp}), (\ref{eq:asymmetric3}) and (\ref{eq:asymmetric4}) to obtain
  $X_{\indHB-1} = A_{\indHB-1} + B_{\indHB-1} = G^{-1}\Af_{\indHB} + G \Bf_{\indHB}$ and 
  $X_{\indHB+1} = A_{\indHB+1} + B_{\indHB+1} = H \Ar_{\indHB} + H^{-1} \Br_{\indHB}.$
Substituting these equations into (\ref{eq:app_hard_boundary1}) and using (\ref{eq:asymmetric2}) for $X_{\indHB}$ yields
\begin{align}
  \Af_{\indHB}(1\!-\!T_{\text{L}} G^{-1})\!+\! \Bf_{\indHB}(1\!-\!T_{\text{L}} G)\! &=\! \Br_{\indHB}(T_{\text{R}}H^{-1})\! +\! \Ar_{\indHB} T_{\text{R}}H.
  \label{eq:app_hard_boundary7}
\end{align}\normalsize
The four wave components are now reduced to three components by substituting $\Ar_{\indHB} = \Af_{\indHB} + \Bf_{\indHB} - \Br_{\indHB}$ or by $\Bf_{\indHB} = \Ar_{\indHB} + \Br_{\indHB} - \Af_{\indHB}$  into (\ref{eq:app_hard_boundary7}). These substitutions give
\begin{align}
  \Bf_{\indHB} &= \Br_{\indHB} \frac{T_{\text{R}} H -T_{\text{R}} H^{-1}}{T_{\text{L}} G + T_{\text{R}}H-1} + \Af_{\indHB}\frac{1-T_{\text{L}} G^{-1}-T_{\text{R}}H}{T_{\text{L}}G + T_{\text{R}}H-1},
  \label{eq:app_hard_boundary8}\\
  \Ar_{\indHB} &= \Af_{\indHB}\frac{T_{\text{L}}G-T_{\text{L}} G^{-1}}{T_{\text{L}} G + T_{\text{R}} H-1} +\Br_{\indHB}\frac{1-T_{\text{L}} G-T_{\text{R}} H^{-1}}{T_{\text{L}} G + T_{\text{R}} H-1}.\label{eq:app_hard_boundary9}
\end{align}
These formulas can be further simplified by expressing $T_{\text{L}}$ and $T_{\text{R}}$ in terms of $G$ and $H$. Specifically, $\olf_{\indHB} = (G + G^{-1}-2)^{-1}$ and $\olr_{\indHB} = (H + H^{-1}-2)^{-1}$. Substituting for them
into (\ref{eq:app_hard_boundary8}) and (\ref{eq:app_hard_boundary9}) yields the result. \qed
\end{pf} 
The theorem states that the wave incident from the left side of the hard boundary (described by $\Af_{\indHB}$) is partially reflected from the boundary (described by $T_{\text{AB}}$) and partially transmitted through the boundary (by $T_{\text{AA}}$). For the wave incident to the opposite side (described by $\Br_{\indHB}$), the transfer functions are $T_{\text{BA}}$ and $T_{\text{BB}}$, respectively. The output of the hard-boundary agent can be expressed in two equivalent ways,
\begin{align}
  X_{\indHB} &=  G (1+T_{\text{AB}}) A_{\indHB-1} + H T_{\text{BB}} B_{\indHB+1},\label{eq:HB_4a}\\
  X_{\indHB} &= G T_{\text{AA}} A_{\indHB-1} + H (1+T_{\text{BA}}) B_{\indHB+1}.\label{eq:HB_4b}
\end{align}
The free-end boundary from (\ref{eq:free_end}) is an example of the hard boundary. In this case $H=0$, which gives $T_{\text{AB}} = G$. Again, if $G(s)=H(s)$, then $T_{AB}=T_{BA}=0$ and $T_{AA}=T_{BB}=1$ which converts (\ref{eq:HB_4a}), (\ref{eq:HB_4b}) into (\ref{eq:pos_decomp}).

\subsection{Properties of the boundaries}
Although the above descriptions of the boundaries are different, they have some common features.  The following result is a direct application of Theorems~\ref{soft_boundary_theorem} and \ref{hard_boundary_theorem}. 
\begin{cor}
\label{cor:HB_tf_relation}
  The soft and hard BTFs are related as follows,
  \begin{align}
    &T_{\textnormal{aa}}(s) = T_{\textnormal{BB}}(s)+G(s)-1,\;\;\;
  T_{\textnormal{ba}}(s) = H(s) T_{\textnormal{AB}}(s),\\
  &T_{\textnormal{bb}}(s) = T_{\textnormal{AA}}(s)+H(s)-1,\;\;\;
  T_{\textnormal{ab}}(s) = G(s) T_{\textnormal{BA}}(s),\\
    &T_{\textnormal{AA}}(s) = 1 + T_{\textnormal{AB}}(s), \;\;\;\;\;\;\;\;\;\;\;\;
    T_{\textnormal{BB}}(s) = 1 + T_{\textnormal{BA}}(s),\label{eq:HB_tf_rel3}\\
    &T_{\textnormal{AA}}(s) + T_{\textnormal{BB}}(s) = 2,\;\;\;\;\;\;\;\;\;\;\;\;
    T_{\textnormal{AB}}(s) + T_{\textnormal{BA}}(s) = 0.\label{eq:HB_tf_rel5}
  \end{align}
\end{cor}

We define the DC gain $\kappa_{\textnormal{G}}$ of $G(s)$ as $\kappa_{\textnormal{G}} = \lim_{s \rightarrow 0} G(s)$. Recall that $\numInteg$ is the number of integrators in the open loop.
\begin{cor}
\label{cor:SB_tf_DC_gain}
  Let $\kappa_{\textnormal{aa}}$ be the DC gain of $T_{\textnormal{aa}}$, $\kappa_{\textnormal{ab}}$ be the DC gain of $T_{\textnormal{ab}}$ etc. If $\numInteg \geq 1$ for both front and rear OLTFs, then 
  \begin{align}
  \kappa_{\textnormal{aa}} + \kappa_{\textnormal{bb}} &= 2,\;\;\; 
  \kappa_{\textnormal{ab}} + \kappa_{\textnormal{ba}}= 0, \label{eq:cor_SB_tf_DC_gain2}\\
  \kappa_{\textnormal{aa}} - \kappa_{\textnormal{ab}} &= 1,\;\;\; 
  \kappa_{\textnormal{bb}} - \kappa_{\textnormal{ba}} = 1,
\label{eq:cor_SB_tf_DC_gain4}\\
\kappa_{\textnormal{aa}} = \kappa_{\textnormal{BB}}, \;\;\; \kappa_{\textnormal{ba}} = &\kappa_{\textnormal{AB}}, \;\;\; \kappa_{\textnormal{bb}} = \kappa_{\textnormal{AA}}, \;\;\; \kappa_{\textnormal{ab}} = \kappa_{\textnormal{BA}},
\end{align}
\end{cor}
\begin{pf}
  Under the above assumptions, the DC gain of a WTF is one, i.e., $\lim G(s)_{s \rightarrow 0} = 1$ and  $\lim H(s)_{s \rightarrow 0}=1$ \cite{Martinec2014a}. Then, the proof is a straightforward application of Corollary~\ref{cor:HB_tf_relation}. \qed
\end{pf}
The DC gains of WTFs and BTFs can be used to approximate the value of the transient (see Sec. \ref{sec:local_DC_gains}). The following values of $\kappa_{\text{aa}}$ and $\kappa_{\text{bb}}$ are proved in \ref{appendix_dcgain_sb}.
\begin{lem}
\label{lem:DC_gain_SB}
Let $\nif, \nir$ be the number of integrators in $\olf_{\indSB}$ and $\olr_{\indSB+1}$, respectively. Assume that $\nif\geq 1, \nir \geq 1$. Then the DC gains of the soft boundary are: i) for  $\nif=\nir$,
\begin{align}
	\kaa = 2\left(\sqrt{\dfrac{n_{1,0}d_{2,0}}{n_{2,0} d_{1,0}}} +1\right)^{-1}, \, \kbb=2\left(\sqrt{\dfrac{n_{2,0}d_{1,0}}{n_{1,0} d_{2,0}}} +1\right)^{-1},
	\label{eq:dcgains}
\end{align}
ii) for $\nif < \nir,$ $\kaa=0, \kbb=2$ \quad and iii) for $\nif>\nir$, $\kaa=2, \kbb=0$. The constants are $n_{1,0}/d_{1,0} = \lim_{s \rightarrow 0} s^\nir \olr_{\indSB}$, $n_{2,0}/d_{2,0} = \lim_{s \rightarrow 0} s^\nif \olf_{\indSB+1}$.
  \end{lem}

\begin{cor}
\label{lem:DC_gains_bounds}
If $\numInteg \geq 1$ both in front OLTF and rear OLTF, then the DC gains are bounded as
  \begin{align}
   -1 &\leq \kappa_{\textnormal{ab}}, \kappa_{\textnormal{ba}}, \kappa_{\textnormal{AB}}, \kappa_{\textnormal{BA}} \leq 1,\\
   0 &\leq \kappa_{\textnormal{aa}}, \kappa_{\textnormal{bb}}, \kappa_{\textnormal{AA}}, \kappa_{\textnormal{BB}} \leq 2.
  \end{align}
\end{cor}
\begin{pf}
The soft-boundary DC gains come from Lemma~\ref{lem:DC_gain_SB} and Corollary~\ref{cor:SB_tf_DC_gain}, the hard-boundary DC gains from Corollary~\ref{cor:SB_tf_DC_gain}.\qed
\end{pf}

\section{Controllers for the boundaries}
We now design a feedback controller compensating the fact that the agents are not identical. The motivation is to prevent the undesired reflection of the wave to shorten the settling time. To control the wave at the boundary, we first have to assume that the external input in (\ref{eq:eq1}) is nonzero and investigate its properties.
For now, we consider that $\inpr_{i}\neq 0$ and we set $\inpf_{i}=0$. Hence, (\ref{eq:eq1}) for the $i$th agent changes to
\begin{align}
  X_i(s)\!&=\! \olf_{i}(s) \left(X_{i-1}(s)\!-\!X_i(s)\right)\! +\!\olr_{i}(s) \left(X_{i+1}(s)\!-\!X_i(s)\!+\!\inpr_i(s)\right).\label{eq:add_input_1}
\end{align}
The input $\inpr_i(s)$ generates a wave that propagates in the system in the same manner as (\ref{eq:anp1}) and (\ref{eq:bnp1}), only (\ref{eq:waveProp}) is changed as follows
\begin{align}
  X_i(s) = G(s)A_{i-1}(s) + G(s)B_{i+1}(s) + \extTr_i (s) \inpr_i(s),
\end{align}
where $\extTr(s) = X_i(s)/ \inpr_i(s)$ for $N \rightarrow \infty$. Since $N \rightarrow \infty$, $X_{i-1}=G X_i, X_{i+1}=G X_i$ and using $X_i = \extTr_i \inpr_i$, we get from (\ref{eq:add_input_1})
\begin{align}
  \extTr_i(s) = \frac{\olr_i(s)}{1 + \Big(1-G(s)\Big) \left(\olf_i(s) + \olr_{i}(s)\right)}.\label{eq:add_input_4}
\end{align}
Analogously, we can calculate the transfer function $\extTf_i(s) = X_{i}(s)/\inpf_{i}$ for a non-zero input $\inpf_{i}$.

\subsection{Soft boundary controller}
\label{subsec:control_soft_boundary}
A soft-boundary controller can be designed for various purposes, for instance, to prevent or modify a wave's transmission through the boundary. We now design an absorbing controller that prevents the reflection of a wave by specifying $\inpf$ and $\inpr$.

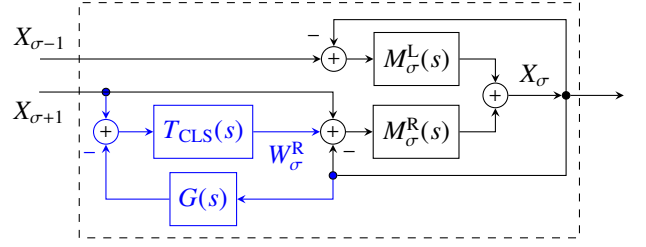
\begin{figure}[t]
 \centering
  \begin{tikzpicture}[auto, >=stealth]
	\node[sum] (sum1) {\footnotesize $+$};
	\node [sum, fill=black, minimum width=0.01em, minimum height=0.01em] (p1) [right= 2em of sum1] {}; 
	\draw [->] (sum1) -- (p1) node[midway, above] { $X_\indSB$};
	\draw [->] (p1.east) -- ([xshift=2em]p1.east);
	
	\node[block] (Cf) [above left= 0.01em and 1em of sum1] {$\olf_{\indSB}(s)$};
	\draw [->] (Cf) -| (sum1);
	\node[block] (Cr) [below left= 0.01em and 1em of sum1] {$\olr_{\indSB}(s)$};
	\draw [->] (Cr) -| (sum1);
	\node[sum] (sumr1) [left= 1em of Cr] {\footnotesize{$+$}}
		edge [->] (Cr);
	\node[sum] (sumf1) [left= 1em of Cf] {\footnotesize{$+$}}
		edge [->] (Cf);
	\node[block, draw=blue] (T) [left = 2.5em of sumr1] { \textcolor{blue}{$T_{\mathrm{CLS}}(s)$}};
	\draw [->, blue] (T) -- (sumr1) node[midway, below] {\textcolor{blue}{$\inpr_\indSB$}};

	\node [sum, fill=blue, minimum width=0.01em, minimum height=0.01em] (p2) [below= 1em of sumr1] {}; 
	\draw [->] (p2) -- (sumr1) node [near end, right] {\footnotesize $-$};
	\draw [] (p1) |- (p2);
	\node [block, draw=blue] (G) [below = 0.5em of T] { \textcolor{blue}{$G(s)$}};
	\draw [<-, blue] (G) -| (p2);
	
	\node [sum, draw=blue] (sum2) [left=8.5em of sumr1.center, anchor=center] {\footnotesize $+$}
		edge [->, blue] (T);
	\draw [->, blue] (G) -| (sum2) node[at end, left, yshift=-1mm] {\footnotesize $-$};
	\node [sum, fill=blue, minimum width=0.01em, minimum height=0.01em] (p3) [above= 1.5em of sum2.center, anchor=center] {}
		edge [->, blue] (sum2);
		 
	\draw [->] ([xshift=-11em]sumf1.center) -- node[at start, above,name=p4] {$X_{\indSB-1}$} (sumf1.west);
	\draw [->] ([xshift=-2.5em]p3.center) -| node[at start, below] {$X_{\indSB+1}$} (sumr1);
	%
	
	\draw[<-] (sumf1.north) -- ++(0.0em,1em)node[near start, xshift=-0.15em, yshift=+0.15em] {\footnotesize$-$} -| (p1.south) ;
	
	\draw [dashed] ([yshift=-5.5em, xshift=-1em]p3) rectangle ([yshift=3.5em, xshift=0.5em]p1);
\end{tikzpicture}
  \caption{The model of $\indSB$th agent with the left-side absorbing controller (blue), where $T_{\text{CLS}}(s) = G(s) (G(s)-H(s))/(1-G^2(s))$ from (\ref{eq:control_soft11a}).}
  \label{fig:soft_controller_model}
\end{figure}

\begin{thm}
\label{lem:soft_boundary_control}
  Suppose that the soft boundary is between agents $\indSB$ and $\indSB+1$. Then the control law that prevents a wave to be reflected from the soft boundary is
  \begin{align}
    \inpr_{\indSB}(s) &= \frac{G(s)\big( G(s) - H(s) \big)}{1-G^2(s)} \big(X_{\indSB-1}(s) - G(s) X_{\indSB}(s) \big),\label{eq:control_soft11a}\\
    \inpf_{\indSB+1}(s) &= \frac{H(s)(H(s)-G(s))}{1-H^2(s)} \left(X_{\indSB+2}(s) - H(s) X_{\indSB+1}(s) \right) \label{eq:control_soft11b}
  \end{align}
\end{thm}
\begin{proof}
We show here how to prevent the reflection only of a wave travelling from the left. Combining (\ref{eq:SB_4a}) and (\ref{eq:add_input_4}), we get
\begin{align}
  X_{\indSB} = (1+T_{\text{ab}}) A_{\indSB} + T_{\text{bb}} B_{\indSB+1} + \extTr_\indSB (1+T_{\text{ab}}) \extInputSB. \label{eq:control_soft3}
\end{align}
The reflection is described by the term $T_{\text{ab}} A_{\indSB}$. Therefore, we eliminate  $T_{\text{ab}} A_{\indSB}$ by requiring
   $T_{\text{ab}}A_{\indSB} + \extTr_\indSB (1+T_{\text{ab}}) \inpr_\indSB = 0,$ 
from which we calculate the external input 
  $\inpr_\indSB = - A_\indSB T_{\text{ab}} / \Big(\extTr_\indSB + \extTr_\indSB T_{\text{ab}}\Big)$.
Substituting for $T_{\text{ab}}$ from (\ref{eq:sb_2}) and for $\extTr_\indSB$ from (\ref{eq:add_input_4}), $\inpr_\indSB$ simplifies to
 \begin{equation} 
 	\inpr_\indSB(s) = \Big(G(s) - H(s)\Big)A_\indSB(s). \label{eq:extInpSB}
 \end{equation}
It remains to specify $A_{\indSB}$. By (\ref{eq:pos_decomp})-(\ref{eq:waveProp}), we have
  $A_{\indSB} = G (X_{\indSB-1} - B_{\indSB-1}) = G X_{\indSB-1} - G^2(X_{\indSB}-A_{\indSB})$, so we can write
\vspace{-5pt}
\begin{align}
   A_{\indSB}(s) &= \frac{G(s)}{1-G^2(s)} (X_{\indSB-1}(s) - G X_{\indSB}(s)).
  \label{eq:control_soft5}
\end{align}
	Plugging (\ref{eq:control_soft5}) to (\ref{eq:extInpSB}) yields (\ref{eq:control_soft11a}). The derivation of control law preventing reflection of the wave travelling from the right---equation (\ref{eq:control_soft11b})---is analogous. 
\end{proof}
Note that the soft-boundary absorbing control requires to control both agents $\indSB$ and $\indSB+1$. The $\indSB$th agent with implemented absorbing controller for the wave from the left is in Fig.~\ref{fig:soft_controller_model}.
Having both absorbers implemented, by modifying (\ref{eq:control_soft3}) we calculate the outputs of the agents $\indSB$, $\indSB+1$ as
\begin{align}
  X_{\indSB} =& (1+T_{\text{ab}}) A_{\indSB} + T_{\text{bb}} B_{\indSB+1} + \extTr_\indSB (1+T_{\text{ab}}) \inpr_\indSB \nonumber\\
  &+ T_{\text{bb}} \extTf_{\indSB+1} \inpf_{\indSB+1} = G A_{\indSB-1} + G B_{\indSB+1},
  \label{eq:control_soft8} \\
  X_{\indSB+1} =& H A_{\indSB} + HB_{\indSB+2}.\label{eq:control_soft9}
\end{align}
Both correspond to a wave propagation in homogeneous system (\ref{eq:waveProp}).
Theorem~\ref{lem:soft_boundary_control} describes the only control law that fully absorbs the wave ($T_{\text{ab}} A_\indSB$ in (\ref{eq:control_soft3}) was cancelled by the unique $\extInputSB$). 

\subsection{The hard boundary controller}
Controlling the hard boundary is similar to that of the soft boundary. Here, we only provide a brief description of the absorbing-controller design.
The output of the $\indHB$th agent from (\ref{eq:HB_4a}) controlled with additional input $\inpf_\indHB(s)$ is
\begin{align}
  X_{\indHB} = (1 + T_{\text{AB}}) \Af_{\indHB} + T_{\text{BB}}\Br_{\indHB} +  \extTf_\indHB \inpf_\indHB,
\end{align}
To prevent the reflection of the wave travelling towards the hard boundary from the left, we set $T_{\text{AB}} \Af_\indHB =- \extTf_\indHB \inpf_\indHB$. Hence,
\begin{align}
  \inpf_\indHB = -\frac{T_{\text{AB}}}{\extTf_\indHB}\Af_\indHB = \frac{(H -G)(1-G)}{G(1-H)}\Af_\indHB.
\end{align}
The $\Af_{\indHB}$ term represents the wave travelling towards the hard boundary from left, which is again computed by (\ref{eq:control_soft5}).


\begin{thm}
\label{lem:hard_boundary_control}
  Suppose that the hard boundary is at the agent $\indHB$. Then the control law that prevents a wave to be reflected from the hard boundary is given in the Laplace domain as
\begin{align}
  \inpf_\indHB(s) &= \frac{H(s)-G(s)}{(1+G(s)) (1-H(s))}( X_{\indHB-1}(s) - G(s) X_{\indHB}(s)), \label{eq:clh} \\
  \inpr_\indHB(s)  &= \frac{G(s)-H(s)}{(1+H(s)) (1-G(s))}( X_{\indHB+1}(s) - H(s) X_{\indHB}(s)).\label{eq:crh}
\end{align} 
\end{thm}
Note that the absorbing controller for the hard boundary is implemented at a single agent. 
The control laws implemented on the agent $\indHB$ using (\ref{eq:HB_tf_rel3}) give
\begin{align}
 X_{\indHB}(s) &= \Af_{\indHB}(s) + \Br_{\indHB}(s) = G(s) A_{\indHB-1}(s) + H(s) B_{\indHB+1}(s), \label{eq:control_hard9}
\end{align}
Combining (\ref{eq:control_hard9}) and (\ref{eq:asymmetric2}) gives $\Af_{\indHB} = \Ar_{\indHB}$ and $\Bf_{\indHB} = \Br_{\indHB}$. In words, the hard boundary at the $\indHB$th agent between the wave components indexed by $\text{L}$ and $\text{R}$ is removed and the wave transmits through the agent without being reflected.


\subsection{End absorbers}
As we discussed before, a forced-end boundary is a special case of the soft boundary. Therefore, setting $G=0$ in (\ref{eq:control_soft11a}), plugging such external input to (\ref{eq:eq1}) and simplifying yields
\begin{align}
  \inpr_1(s) &= G(s)\big[X_1(s) - G(s) \posRef(s)].\label{eq:wtf_intro5}
\end{align}
Similarly, since the free-end boundary is a special case of the hard boundary, setting $H=0$ in (\ref{eq:control_soft11b}), plugging it to (\ref{eq:eq1}) and separating the part for the external control law gives
\begin{equation}
	\inpf_N(s) = \big(G(s)-1\big) \pos_N(s) \label{eq:free_end_absorber}
\end{equation}
Cancelling the wave reflections at these boundaries shortens the transient time even qualitatively \cite{Martinec2014a}.

\subsection{Stability of the controllers}

To prove stability of the control law, we first need to prove stability of the WTF given some assumptions on the open-loop transfer function $\ol(s)$. The proof is given in \ref{appendim_stability_wtf}.
\begin{lem}
\label{lem:stability_wtf}
If $M(s)$ is proper and has no CRHP (closed right half-plane) zeros and no CRHP poles, except for poles at the origin, and if the Nyquist plot of 
  $M(\jmath \omega)$
does not intersect the real interval $(-\infty, -1/4)$, then the wave transfer function $G(s)$ in (\ref{eq:wtf_intro1}) corresponding to $M(s)$ is asymptotically stable.
\end{lem}

We show below that the controllers do not destabilize the system (proof is in \ref{appendix_soft_boundary_stability}). The opposite is true: their use make the system even string stable for arbitrary platoon size. This guarantees very good transients.
\begin{defn}[{\cite{Eyre1998a}}]
A system is called $L_2$ string stable if there is an upper bound on the $L_2$-induced system norm of $T_{0,i}(s)$ that does not depend on the number of agents, where $T_{0,i}(s)=\frac{X_i(s)}{X_{\text{ref}}(s)}$.\label{def:string_stability}
  \end{defn}
\begin{thm}
\label{lem:soft_boundary_stability}
If the WTFs are asymptotically stable, then the multi-agent system with the path-graph topology and the control laws from Theorem~\ref{lem:soft_boundary_control} or Theorem~\ref{lem:hard_boundary_control} is asymptotically stable. Furthermore, these control laws implemented together with the end-node wave absorbers, located on the \nth{1} or $N^{\mathrm{th}}$ agent, make the multi-agent system $L_2$ string stable.
\end{thm}
In order to guarantee stability, all WTFs corresponding to the open loops of the agents must satisfy Lemma \ref{lem:stability_wtf}. This, combined with boundary controllers, gives a sufficient stability condition. Note that without the boundary absorbers, it is hard to make any stability statement using wave approach.

\subsection{Discussion of wave controllers}
The advantage of the wave controller is that it allows the modification of the reflection conditions for the travelling waves on a boundary. Importantly, this modification does not require to change controllers of the other agents in the system and, under certain conditions, it can make the system string stable. Another advantage is that the output of the controller is feasible (see Fig.~\ref{system_input_comp_SB}). A difficulty is that the agent is required to know its own and neighbour's dynamical models. If the neighbour's model is known only approximately, then the wave is not fully absorbed and it partially reflects back. However, the numerical simulations show that the response of the system may still be improved since absorbers are relatively `robust' to the inaccuracies. This is in agreement with experience from practical implementations, see for instance \cite{Saigo2004}, \cite{Kreuzer2011} or \cite{OConnor2008a}.
\vspace{-10pt}
\section{Numerical simulations}
\begin{figure*}[ht]
 \centering
  \includegraphics[width=0.95\textwidth]{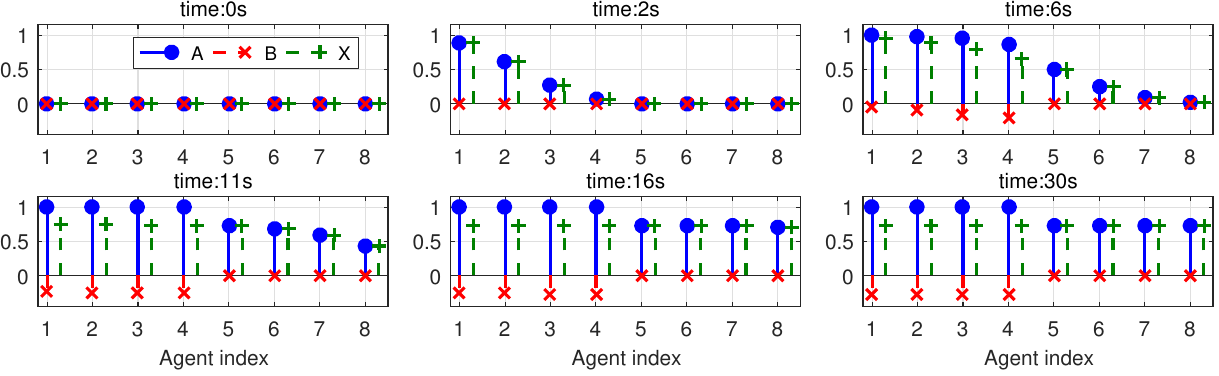}
  \caption{Simulation of the wave propagating in a system with $N=8$. The \nth{1} agent implements the forced-end absorbing control law (\ref{eq:wtf_intro5}) and the \nth{8} agent implements free-end absorber (\ref{eq:free_end_absorber}). A response to a unit-step change in $X_\mathrm{ref}$ is shown. At intermediate times, the wave travels to the soft boundary between agents 4 and 5, where it is transmitted and attenuated by a factor $\kappa_{\text{aa}}=0.73$ (from (\ref{eq:dcgains})) and reflected by a factor $\kappa_{\text{ab}}=-0.27$ (from (\ref{eq:cor_SB_tf_DC_gain4})). As the partially reflected wave propagates back to the first agent, it forces the first four agents to decrease their output by $\kappa_{\text{ab}}$.  }
  \label{fig:comp_soft_bound_N10_J}
\end{figure*}
\label{subsec:Numerical_simulations_SB}
The numerical simulations are carried out with our library in MATLAB called \emph{WaveBox} (available at \cite{WaveBox2015}). The \emph{WaveBox} also contains examples that show the effect of boundaries and absorbers. Some of the examples are presented in this section.

In the numerical simulations we use the following OLTFs   
\begin{equation}
\begin{aligned}
  M_1(s) &= \frac{4s+4}{s^2(s+4)},\,  M_2(s)=\frac{s+\gamma}{s^2(s+3)}, \\ 
  M_3(s) & =\frac{4s+4}{s^2(s+3)}, \,M_4(s)=\frac{s+1}{s^2(s+4)} \label{eq:agentModelsSim}
\end{aligned}
\end{equation}
which represent double integrator agents with viscous friction, controlled by PI controllers.

First, consider a path graph of 8 agents with OLTFs $\olf_{i}(s)=\olr_{i}(s)=\ol_1(s)$ for $i=1,2,3,4$ and $\olf_{i}(s)=\olr_{i}(s)=\ol_2(s)$ for $i=5,6,7,8$ and set $\gamma=1$. There is a soft boundary located between the \nth{4} and \nth{5} agent. The effect of the soft boundary along with wave-absorbing controllers on the first and last agents is demonstrated in Fig.~\ref{fig:comp_soft_bound_N10_J}. The reflection at the boundary is revealed by the negative value of the $B$ component at the end.


For a system with multiple boundaries (described in Fig. \ref{fig:multBoundDesc}), in Fig. \ref{fig:Six_figs_pos_SB} we evaluate the transient responses for various absorbers. Comparing the figures in the left column, we see that implementing the boundary absorbers improves the transient time. The reason is that without boundary absorbers, the reflections at the boundaries attenuate the wave propagating to the end, so the agents at the end react more slowly. For the absorber on the first agent (top-middle panel), the wave keeps reflecting between the boundaries, which prolongs the transient. The implementation of the boundary absorber (bottom-middle panel) shortens the transient a lot. The absorbers implemented on both the first and last agents (top-right panel) cause a change of the steady-state value. This is solved by adding boundary absorbers to all boundaries (bottom-right panel). Note that this solution shortens the transient time more than 10 times and removes oscillations from the transient response.


\begin{figure}[t]
 \centering 
  \begin{tikzpicture}[auto, >=latex]
	\node[blueagent]  (N1) {1};	
	\node[blueagent]  (N2)	[right=of N1, right=1.5em] {2}; 	
	\node[placeholder]  (P1)	[right=of N2.center, right=1.1em] {};
	\node[redagent]  (N3)	[right=of P1.center, right=0.8em] {3}; 
	\node[agent, shading = axis,left color=red!80, right color=green!80]  (N4)	[right=of N3, right=1.5em] {4}; 
	\node[greenagent]  (N5)	[right=of N4, right=1.5em] {5}; 
	\node [placeholder] (P2) [right=of N5.center, right=1.1em] {};
	\node[orangeagent]  (N6)	[right=of P2.center, right=0.8em] {6}; 
	\node[agent, shading = axis,left color=orange!80, right color=blue!80]  (N7)	[right=of N6, right=1.5em] {7}; 
	\node[blueagent]  (N8)	[right=of N7, right=1.5em] {8};
	\draw [bluespringShort] (N1) -- (N2);
	\draw [line width=1pt,decorate,decoration={zigzag,pre length=0.05cm,segment length=3}, draw=blue!80] (N2) -- (P1.center);
	\draw [line width=1pt,decorate,decoration={zigzag,pre length=0.05cm,post length=0.0cm, segment length=3}, draw=red!80] (N3) -- (P1.center);
	\draw [redspringShort] (N3) -- (N4);
	\draw [greenspringShort] (N4) -- (N5);
	\draw [line width=1pt,decorate,decoration={zigzag,pre length=0.05cm,segment length=3}, draw=green!80] (N5) -- (P2.center);
	\draw [line width=1pt,decorate,decoration={zigzag,pre length=0.05cm,post length=0.0cm, segment length=3}, draw=orange!80] (N6) -- (P2.center);
	\draw [orangespringShort] (N6) -- (N7);
	\draw [bluespringShort] (N7) -- (N8);

\end{tikzpicture}
  \caption{Schematic of the simulation setup for $N=8$. There are two soft boundaries (between $2,3$ and between $5, 6$) and two hard boundaries (at agents $4, 7$). Blue color corresponds to $M_1$, red to $M_2$, green to $M_3$ and orange to $M_4$.} 
  \label{fig:multBoundDesc}
\end{figure}
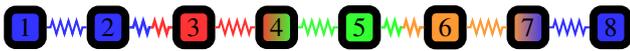
\begin{figure*}[t]
 \centering 
  \includegraphics[width=0.95\textwidth]{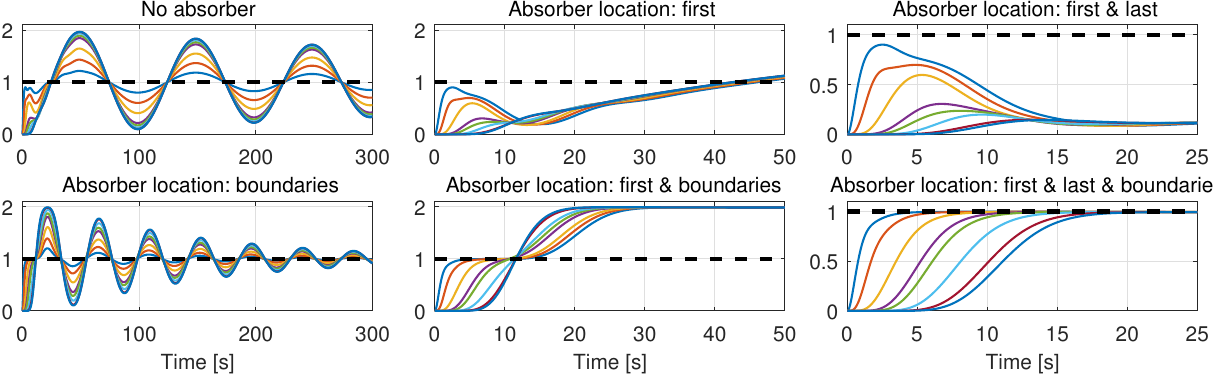}
  \caption{Performance comparison of individual control strategies for a system with multiple boundaries in Fig. \ref{fig:multBoundDesc}. At time 0 all outputs are zero and at time 0 there is a unit step change in the reference $\posRef$ for the first agent. Note different time scales of the columns. Dashed line is $\posRef=1$.}
  \label{fig:Six_figs_pos_SB}
\end{figure*}
 
\begin{figure}[t] 
 \centering
  \includegraphics[width=0.49\textwidth]{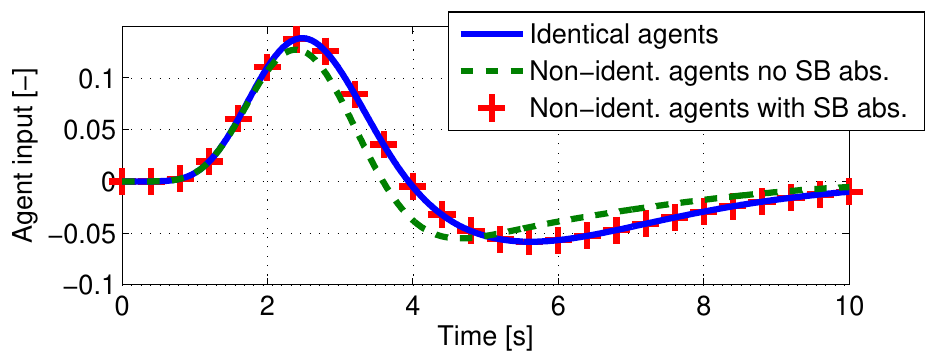}
  \caption{The comparison of the inputs to the fourth agent for three different multi-agent systems. The label 'SB abs.' stands for the soft-boundary absorber.}
  \label{system_input_comp_SB}
\end{figure}

Fig.~\ref{system_input_comp_SB} shows the comparison of the inputs to the fourth agent for (i) the homogeneous system with 8 identical agents (blue solid line) with $\olf_{i}(s)=\olr_{i}(s)=M_1(s)$, (ii) the system with non-identical agents as in Fig. \ref{fig:comp_soft_bound_N10_J} without the soft-boundary absorber (green dashed line), and (iii) the system as in (ii) but with the soft-boundary absorber (red pluses). It illustrates that the absorber does not change the control effort too much.

\subsection{Local effect of the DC gains}\label{sec:local_DC_gains}

\begin{figure}[t]
 \centering
  \includegraphics[width=0.40\textwidth]{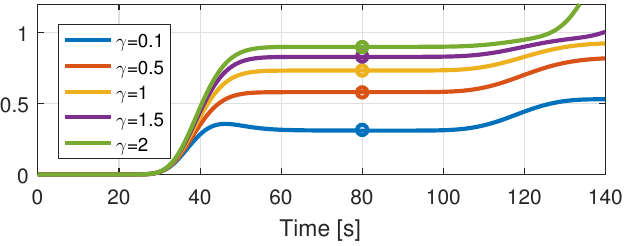}
  \caption{Output of the $41^{\text{st}}$ agent with $N=80$ for different values of $\gamma$. The circles show the value of the DC gain of $T_{\text{aa}}$ from (\ref{eq:dcgains}). The approximation is therefore precise for $t \in [50, 100]$.}
  \label{fig:DC_gains_comp}
\end{figure}

A local effect of the BTF DC gains is demonstrated in Fig.~\ref{fig:DC_gains_comp}. It shows the first 140 seconds of the step response of system with $N=80$. The models are $\olf_{i}(s)=\olr_{i}(s)=\ol_1(s)$ for $i=1,\ldots,40$ and $\olf_{i}(s)=\olr_{i}(s)=\ol_2(s)$ for $i=41, \ldots, 80$. We change the term $\gamma$ in (\ref{eq:agentModelsSim}) to see its effect. Applying (\ref{eq:pos_decomp})-(\ref{eq:free_end}) and (\ref{eq:SB_4b}), the output of the \nth{41} agent is 
\begin{align}
  X_{41}(s) = T_{\text{aa}}(s) A_{40}(s) + (1+T_{\text{ba}}(s))B_{41}(s),
\end{align}
where $B_{41}(s)$ is the wave that first transmitted through the boundary, travelled from the \nth{41} agent to the \nth{80} agent, reflected at the end and then travelled back to the \nth{41} agent, hence, $B_{41}(s) = T_{\text{aa}}(s) H^{79}(s) A_{40}(s)$. Before this reflected wave reaches the \nth{41} agent ($t \approx 110$), we can approximate the output $X_{41}(s)$ as $X_{41}(s) \approx T_{\text{aa}}(s) A_{40}(s)$. After the initial settling time, this approximation is the DC gain of $T_{\text{aa}}(s)$, so
  $x_{41}(t) \approx \kappa_{\text{aa}}, \, t \in [50, 100]$. Hence, for some time the output of the \nth{41} agent equals the DC gain of $T_{\text{aa}}$ (\ref{eq:dcgains}). The most important feature of the approximation is that it does not consider interactions among other agents. In other words, there can be arbitrary number of agents with various dynamics after the \nth{41} agent. 

\section{Conclusion}  


This paper introduces a wave approach to a multi-agent system with a path-graph topology and heterogeneous agents. We define and mathematically describe two basic types of boundaries between non-identical agents. Their effect on the waves travelling in the system is captured using transfer functions, which show that part of the wave gets through the boundary and part is reflected back. The travelling waves describe the system from a local point of view. 
 
 The wave description allows us to design a feedback controller to compensate the effect of the boundary, which shortens the settling time. Moreover, such a controller makes the system string stable provided that the system is equipped with at least one wave absorber on the first or last agents.
 
 The irrational transfer functions allow easy local description: they reveal how the waves reflect, what is the steady-state gain before the waves reflect back or whether the wave is amplified. Using wave absorbers we can also significantly shorten the transient. Recent result \cite{Martinec2015b} shows that the wave-based approach is also useful as an analytical tool also even in cases without absorbers---to describe a local amplification of the travelling wave. On the other hand, unless the wave absorbers are implemented, using the wave perspective it is very difficult to test for instance stability of the heterogeneous system or calculate its $\mathcal{H}_\infty$ norm. This is where the traditional state-space methods are more useful. Thus, we consider the wave-based approach as a complementary tool to standard state-space approaches.  

\appendix

\vspace{-5pt}
\section{Proof of Lemma~\ref{lem:DC_gain_SB}}
\label{appendix_dcgain_sb}
Let us denote $\alr = 2+1/\olr_{\indSB} = 2+d_1/n_1 $ and $\alf = 2+1/\olf_{\indSB+1} = 2+d_2/n_2$. We will show the DC gain $\kaa$ of the $T_{\text{aa}}$ transfer function. The derivation of $\kbb$ is similar.
\begin{align}
  \kaa &= \lim_{s \rightarrow 0} T_{\text{aa}} = \lim_{s \rightarrow 0} \frac{H-H G^2}{1-HG} = \lim_{s \rightarrow 0} \frac{1-G^2}{H^{-1}-G} = \frac{0}{0},\label{eq:proofA_0}
\end{align}
since $\lim_{s \rightarrow 0} G= \lim_{s \rightarrow 0} H = 1$ for at least one integrator in $\olr_{\indSB}$ and $\olf_{\indSB+1}$. Applying the L'Hopital's rule to (\ref{eq:proofA_0}) gives
\begin{align}
  \kappa_{\text{aa}} &= \lim_{s \rightarrow 0} \frac{2G}{H^{-2} H'\, (G')^{-1} +1} = \lim_{s \rightarrow 0} \frac{2}{H'\, (G')^{-1} +1},
  \label{eq:proofA_1}  
\end{align}
where the symbol $'$ denotes the differentiation with respect to variable $s$. First, the differentiation of (\ref{eq:sb_3}) yields
  $G' = \frac{1}{2} \alr' - \frac{1}{2}\frac{\alr \alr'}{\sqrt{\alr^2 - 4}}.$
For $\alr'$ we get that if $\nir=1$, then $\lim_{s\rightarrow 0} \alr' < \infty$ and if $\nir>1$, then $\lim_{s\rightarrow 0} \alpha_1' = 0$.
The other part of $G'$ is
\begin{align}
&\lim_{s\rightarrow 0} \frac{\alr'}{\sqrt{\alr^2-4}} = \lim_{s\rightarrow 0} \dfrac{d_1'n_1 -d_1n_1'}{n_1\sqrt{d_1^2+4d_1n_1}} \nonumber\\
&= \lim_{s \rightarrow 0} \dfrac{ \left(s^{(\nir/2)-1} \nir d_{1,0}n_{1} \right)}
{n_1 \sqrt{4 d_{1,0}n_1}} =
\lim_{s \rightarrow 0} \nir s^{(\nir/2)-1} \frac{1}{2} \sqrt{\frac{d_{1,0}}{n_{1,0}}},
\end{align}
where $d_{1,0} = \lim_{s \rightarrow 0} s^{-\nir} d_1$ and $n_{1,0} = \lim_{s \rightarrow 0} n_1$. Moreover $\lim_{s \rightarrow 0} \alr = 2$. Similarly, $\lim_{s \rightarrow 0}H'$ can be evaluated. Then
\begin{align}
  \lim_{s \rightarrow 0} \frac{H'}{G'} = \lim_{s \rightarrow 0} s^{(\nif-\nir)/2} \frac{\nif}{\nir} \sqrt{\frac{n_{1,0}d_{2,0}}{n_{2,0}d_{1,0}}},
  \label{eq:proofA_2}
\end{align}
where $d_{2,0} = \lim_{s \rightarrow 0} s^{-\nif} d_2$ and $n_{2,0} = \lim_{s \rightarrow 0} n_2$.
Finally, substituting (\ref{eq:proofA_2}) into (\ref{eq:proofA_1}) yields
  $\kappa_{\text{aa}} = \lim_{s \rightarrow 0} 2\left(s^{(\nif-\nir)/2} \dfrac{\nif}{\nir} \sqrt{\dfrac{n_{1,0}d_{2,0}}{n_{2,0}d_{1,0}}} + 1\right)^{-1}.$
Assigning the particular values for $\nif, \nir$ gives the result.\qed

\section{Proof of Lemma~\ref{lem:stability_wtf}}
\label{appendim_stability_wtf}

The proof is based on \cite[Thm. A.2]{Curtain2009}, which states: \emph{A linear system is stable if and only if its transfer function $T(s)$ is analytic in the right-half plane (RHP) and $||T||_{\infty} < \infty$, where $||T||_{\infty} = \sup_{Re(s) >0} |T(s)|$.}

It was proved in \cite{Martinec2014a} that WTF in (\ref{eq:wtf_intro1}) satisfies $||G||_{\infty} \leq 1$. Hence, here we only show that WTF is analytic in RHP.

First, we treat the square root function in (\ref{eq:wtf_intro1}). By \cite[pp. 99-100]{stein2010complex},  $f(z)=\sqrt{z}$ is analytic everywhere, except for the non-positive real axis. Therefore, $f_2(\alpha)=\sqrt{\alpha^2-4}$ in (\ref{eq:wtf_intro1}) is analytic for $\alpha \notin [-2,2]$. Since $\alpha(s) = 2+1/M(s)$, $M(s)$ must not be equal to $(-\infty,-1/4]$ for $\Re\{s\}>0$. Since by assumption $M(s)$ has no CRHP pole except for $\numInteg$ poles at origin, we can invoke the Maximum modulus principle \cite[Thm. 4.5 ]{stein2010complex} and test the condition $\ol(s) \neq (-\infty,-1/4]$ for $\Re\{s\}>0$ by the condition $\ol(\jmath \omega) \neq (-\infty,-1/4]$ for $\omega \in \mathbb{R}$---that is, by the Nyquist plot of $M(s)$. It follows that to guarantee analyticity, $M(\jmath \omega)$ must not not intersect the real interval $(-\infty,-1/4]$ for $\omega \in (0,\infty)$.

The other part of the $G(s)$, $1 + 0.5/M(s)$, is analytic in the CRHP if and only if there are no CRHP zeros of $M(s)$. If both parts of $G(s)$ are analytic, their difference is also analytic and $G(s)$ is a stable transfer function. \qed

\vspace{-10pt}
\section{Proof of Theorem~\ref{lem:soft_boundary_stability}}
\label{appendix_soft_boundary_stability}
Consider a system with one soft boundary (between $\indSB$ and $\indSB+1$) with models $\olf_{i}=\olr_{i}$, $i=1,...,\indSB$ and $\olf_{j}=\olr_{j}$, $j =\indSB+1,..., N$.  The proof for multiple soft or hard boundaries is analogous. We omit the `s' operator from the equations.


First, consider a forced- or free-end wave absorber only at the first agent. Then the combination of (\ref{eq:anp1}), (\ref{eq:bnp1}), (\ref{eq:control_soft8}) and (\ref{eq:control_soft9}) gives
\begin{align}
  \frac{X_p}{X_{\text{ref}}} = \begin{cases}
  	G^p + G^{2\indSB+1-p}H^{2(N-\indSB)} & \text{ if } p\leq \indSB,\\
  	G^\indSB H^{p-\indSB}+G^\indSB H^{2N+1-\indSB-p} & \text{ if } p > \indSB.
  \end{cases}\label{eq:pf_st_01}
\end{align}
In the alternative case of the wave absorbers implemented either on the rear-end agent, or on both the ends, we get
\begin{align}
  \frac{X_p}{X_{\text{ref}}} = \begin{cases} 
  	G^p & \text{ if } p\leq \indSB,\\
  	G^\indSB H^{p-\indSB} & \text{ if } p > \indSB.
  \end{cases}\label{eq:pf_st_03}
\end{align}
Since $G$ and $H$ are asymptotically stable, and $\|G\|_{\infty}\leq 1$ and $\|H\|_{\infty}\leq 1$, then (\ref{eq:pf_st_01}),(\ref{eq:pf_st_03}) are asymptotically stable too and their the $H_{\infty}$ norm is limited regardless of $N$. By Definition~\ref{def:string_stability}, these two systems are also $L_2$ string stable.

Now, we prove stability for no wave absorber on either ends, but with a soft-boundary absorber at agent $\indSB$. Using the properties of the boundaries (\ref{eq:forced_end}, \ref{eq:free_end}) and the soft boundary absorbing control (\ref{eq:control_soft8}-\ref{eq:control_soft9}), we can derive $A_1=G \posRef -G^{2\indSB+1} H^{2(N-\indSB)}A_1$ and $B_1 = G^{2\indSB} H^{2(N-\indSB)} \posRef - G^{2\indSB+1} H^{2(N-\indSB)} B_1$, hence
\begin{align}
   \frac{X_1}{X_{\text{ref}}} &= \frac{A_1+B_1}{X_{\text{ref}}} = \frac{G+G^{2\indSB}H^{2(N-\indSB)}}{1+G^{2\indSB+1}H^{2(N-\indSB)}},
   \\\frac{X_2}{X_{\text{ref}}} &= \frac{A_2+B_2}{X_{\text{ref}}} = \frac{G^2+G^{2\indSB-1}H^{2(N-1\indSB)}}{1+G^{2\indSB+1}H^{2(N-\indSB)}},
\end{align}
and so on. For the $p$th agent, we have
\begin{align}
  \frac{X_p}{X_{\text{ref}}} = \begin{cases}
  \displaystyle{\frac{G^p + G^{2\indSB+1-p} H^{2(N-\indSB)}}{1+G^{2\indSB+1}H^{2(N-\indSB)}}} & \text{ if } p\leq \indSB,\\
  \displaystyle{\frac{G^\indSB H^{p-\indSB} + G^{\indSB} H^{2N-\indSB-p+1}}{1+G^{2\indSB+1}H^{2(N-\indSB)}}} & \text{ if } p > \indSB.
  \end{cases} \label{eq:pf_st_2}
\end{align}
Note that the transfer functions between two arbitrary agents can be expressed similarly.

By (\ref{eq:wtf_intro1}), $M/(1+2M) = G /(1+G^2)$. If $M/(1+\lambda M)$ is stable for $\lambda \in (0,4]$, and if there are neither CRHP poles nor CRHP zeros in $M(s)$, then the Nyquist curve of $M(s)$ does not encircle the point $[-1/4,0]$. Hence, $M(s)$ does not intersect $(-\infty,-1/4]$. Thus, if the conditions in Lemma \ref{lem:stability_wtf} hold, then $M /(1+2M)$ is stable and also $G/(1+G^2)$ is stable. Using $||G||_{\infty}\leq 1$ and $||H||_{\infty}\leq 1$, the transfer function $G/(1+G^2 G^{q_1} H^{q_2})$
is asymptotically stable for $q_1, q_2 \in \mathbb{N}$. Furthermore, since $G$ and $H$ are stable, then the following transfer function is stable for $q_3, q_4 \in \mathbb{N}$ 
\begin{align}
  \frac{G G^{q_3} H^{q_4}}{1+G^2 G^{q_1} H^{q_2}}.\label{eq:pf_st_6}
\end{align}
Comparing (\ref{eq:pf_st_6}) with (\ref{eq:pf_st_2}), we proved asymptotic stability for the case with no absorber at the end and boundary control law by Theorem \ref{lem:soft_boundary_control}. The proof for the control law from Theorem~\ref{lem:hard_boundary_control} can be carried out analogously. The only difference is in different powers of $G$ and $H$ in (\ref{eq:pf_st_01}),(\ref{eq:pf_st_03}) and (\ref{eq:pf_st_2}).\qed 
\vspace{-10pt}



\bibliographystyle{elsart-num-sort}

\bibliography{2014-Waves_in_path_graphs}

\end{document}